\documentclass{elsart}   

\usepackage{graphicx}

\begin{document}
\newcommand{\mycaption}[2]{\begin{center}{\bf Figure \thefigure}\\{#1}\\{\em #2}\end{center}\addtocounter{figure}{1}}
\newcommand{\myauthor}{Ingve Simonsen}
\newcommand{\mytitle}{Measuring Anti-Correlations in the Nordic Electricity Spot
    Market by Wavelets}
\newcommand{\R}[1]{(\ref{#1})}

\begin{frontmatter}
\title{\mytitle}
\author{Ingve Simonsen\thanksref{ingve}}

\thanks[ingve]{Email : ingves@nordita.dk}

\address{
  The Nordic Institute for Theoretical Physics --- NORDITA\\
  Blegdamsvej 17, DK-2100 Copenhagen {\O}\\
  DENMARK }



\begin{abstract}
  We consider the Nordic electricity spot market from mid 1992 to the
  end of year 2000. This market is found to be well approximated by an
  anti-persistent self-affine (mean-reverting) walk. It is
  characterized by a Hurst exponent of $H\simeq 0.41$ over three
  orders of magnitude in time ranging from days to years. We argue
  that in order to see such a good scaling behavior, and to locate
  cross-overs, it is crucial that an analyzing technique is used that
  {\em decouples} scales.  This is in our case achieved by utilizing
  a (multi-scale) wavelet approach. The shortcomings of methods that
  do not decouple scales are illustrated by applying, to the same data
  set, the classic $R/S$- and Fourier techniques, for which scaling
  regimes and/or positions of cross-overs are hard to define.
\end{abstract}


\end{frontmatter}


\section{Introduction}

Over the last ten years, or so, dramatic changes have occurred, and
are still on-going, in the energy sectors of the world. What used to
be well established monopolies in many countries and regions, were
deregulated in such a way that consumers could buy their electricity
from other sources then their local provider. These changes opened up
for competition on the price of electric energy.  Energy exchanges
were created, as a result, as places where such organized transactions
could take place.

One of the first electricity markets in the world to be deregulated
was the Norwegian, that was fully deregulated by 1992. The same year a
Norwegian commodity exchange for electric power was established. Here
one could trade short and long term electricity contracts in addition
to contracts for next-day ($24\,h$) or immediate physical delivery.
This market place, which now also includes the other Nordic
countries, is today known as NordPool --- the Nordic Power
Exchange~\cite{NordPool}.

The next-day electricity market, {\it i.e.}, where contracts for
$24\,h$ delivery are traded, is known as the {\em spot market}.  This
market, administrated by NordPool, is open $24$ hours a day $7$ days a
week all year around, and the (spot) price is fixed for each hour
separately.  In this market, participants ({\it i.e.} the buyers and
sellers) tell the market administrator (NordPool) how much, and to
what price and at what time they want to sell or buy a given amount of
electric power.  From these bid and ask data the administrator creates
a {\em market cross} which sets the spot price for that particular
hour.  The market cross is obtained by forming the cumulative volume
histograms, $V_s(p)$ and $V_b(p)$, that give respectively the total
amount of electric energy that sellers (buyers) want to sell (buy) at
a price higher (lower) then $p$. The spot price, $S$, is then defined
as the price at which $V_s(S)=V_b(S)$ (if such a point exists). Thus,
the spot price is set such that the total volume of sold and bought
electric power is balanced. All sellers asking a price $p\leq S$, as
well as all buyers willing to pay $p\geq S$ will get a transaction
{\em at the} spot price for that particular hour. In all other cases
no transactions will take place. If no market cross can be defined
from the bid and ask data, no transactions will take place and the
spot price for that particular hour remains unset.

Recently, it has been suggested that the electricity spot price
process is a so-called mean-reverting
process~\cite{Weron-2000-A,Weron-2000}. Here, the degree of
mean-reversion was quantified for the Californian spot power market by
measuring its Hurst or roughness exponent by the
$R/S$-analysis~\cite{Hurst-1951,Book:Feder-1988}.  However, in order
to perform this analysis, Weron and
Przyby{\l}owicz~\cite{Weron-2000-A} had to consider the daily average
returns instead of the hourly returns.  This is due to a shortcoming
of the $R/S$-method when analysing multi-scale time series as we will
see explicitly in the discuss below.

In this paper, however, we suggest to use another technique --- a
wavelet based analyzing technique that does not suffer from this
shortcoming. The method is demonstrate on data from NordPool --- the
oldest multi-national power exchange in the world. It is found that
the Nordic electricity power market is anti-persistent
(mean-reverting), and the (Hurst) exponent characterizing the market
is found to be $H=0.41 \pm 0.02$ for time scales ranging from days to
years, where the error indicated is the regression error.

\section{The data set}

The data we analyzed are the official NordPool spot price data (system
price)~\cite{NordPool} collected over the period from May 4, 1992 to
December 31, 2000. They cover more then eight and a half years of
hourly logged data, which corresponds to somewhat more then $75\,000$
data points for the whole data set.



The analyzed data set is depicted in Fig.~\ref{Fig:Price}. From this
figure one easily observes the daily and seasonal trends that are
characteristic of the spot price. Such trends are most likely
attributed to the consumption patterns of the market.  For instance,
in the Californian electricity spot market the high price season is
the summer season~\cite{Weron-2000-A} while for the Nordic power
market it is the winter. These high price seasons coincide with the
peaks in consumption which are different due to the climatic
differences between the Nordic countries and California.


Even though we are talking about the Nordic power market, it should be
realized that during the period from 1992 to 2000 the market has
changed its structure when it comes to participating countries. In the
beginning, the market was a pure Norwegian market, because Norway was
the first Nordic country to deregulate its power sector. Roughly five
years later, during late 1995 and early 1996, Sweden and Finland
deregulated their electricity markets, and became members of what
today is the NordPool system. So, first at this point in time it is
fair to talk about a real Nordic Power exchange. Finally, last year
also Denmark joined in, while there for the moment seems to be no
indication that Island, as the last remaining Nordic country, will do
the same.

\section{Self-affine processes and mean-reversion}

A time-dependent function $p(t)$ is said to be self-affine if
fluctuations on different time scales can be rescaled so that the
original signal $p(t)$ is statistical equivalent to the rescaled
version $\lambda^{-H}p(\lambda t)$ for any positive number
$\lambda$~\cite{Book:Feder-1988}, {\it i.e.},
\begin{eqnarray}
    \label{Eq:Self-affine}
    p(t) \simeq \lambda^{-H}p(\lambda t),
\end{eqnarray}
where $\simeq$ is used to denote statistical equality.  Here $H$ is
the so-called Hurst (or roughness) exponent~\cite{Book:Feder-1988}, an
exponent which quantifies the degree of correlation --- positive or
negative --- in the increments $\Delta p(t)=p(t)-p(0)$. It can be
shown that for a process satisfying Eq.~(\ref{Eq:Self-affine}) the
correlation function $C(t)$ between future, $\Delta p(t)$, and past,
$\Delta p(-t)$, increments is given by~\cite{Book:Feder-1988}
\begin{eqnarray}
    \label{Eq:Correlation-func}
     C(t) = \frac{
                    \left< -\Delta p(-t) \Delta p(t) \right>
                 }{
                    \left<\left[\Delta p(t)\right]^2\right>
                  }
          =   2^{2H-1}-1.                       
\end{eqnarray}
From the above expression one should notice that $C(t)$ is
time-independent, and that for a Brownian motion ($H=1/2$) past and
future increments are uncorrelated ($C(t)=0$) as is well-known.
However, if $H>1/2$ the increments are positively correlated
($C(t)>0$) and for $H<1/2$ they are negatively correlated ($C(t)<0$).
In the former case we say that the process $p(t)$ is {\em persistent}
while in the latter one talks about an {\em anti-persistent} (or
anti-correlated) process. In the economics literature anti-persistency
is known under the name of mean-reversion.

\section{The Average Wavelet Coefficient Method}

The Average Wavelet Coefficient
Method~(AWC)~\cite{Sahimi-1997,Simonsen1998-1} is a method that
utilizes the wavelet
transform~\cite{Book:Daubechies-1992,Book:Percival-2000} in order to
measure the temporal self-affine correlations of a time series, {\it
  i.e.}, a method for measuring its Hurst exponent $H$. This is done
by transforming the time series, $p(t)$, into the wavelet-domain,
${\mathcal W}[p](a,b)$, where $a$ and $b$ represent a scale and
location parameter,
respectively~\cite{Book:Daubechies-1992,Book:Percival-2000}. The
AWC-method consists of, for a given scale $a$, to find a
representative wavelet ``energy'' or amplitude for that particular
scale, and to study its scaling.  This can, for instance, be done by
taking the arithmetic average of $|{\mathcal W}[p](a,b)|$ over all
location parameters $b$ corresponding to one and the same scale $a$.
Thus, one can construct from the wavelet transform of $p(t)$, the
AWC-spectrum $W[p](a)$ (to be defined below) so that it will only
depend on the scale $a$.  If $p(t)$ is a self-affine process
characterized by an exponent $H$, this spectrum should scale
as~\cite{Simonsen1998-1,comment}
\begin{eqnarray}
    \label{Eq:AWC-scaling}
    W[p](a) = \left< \left|{\mathcal W}(a,b)\right|\right>_b 
    \sim a^{H+1/2}.
\end{eqnarray}
Thus, if we plot $W[p](a)$ vs. scale $a$ in a log-log plot, the slope
should be $H+1/2$ if the signal is self-affine. We note that this
method is a multi-scale method in the sense that the behavior at
different scales does not influence each other in any significant way,
{\it i.e.}, the method decouples scales as we will see exemplified
below. This decoupling of scales comes about due to the wavelets being
orthogonal~\cite{Book:Daubechies-1992,Book:Percival-2000}.

\section{Numerical results}     
                               
\begin{figure}[tbp]
    \begin{center}
        \leavevmode
         \includegraphics[scale=0.43]{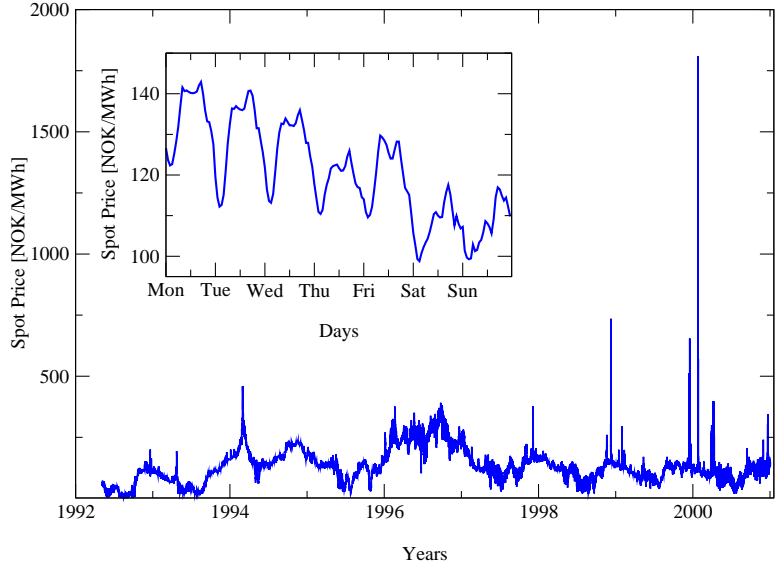}
        \caption{The hourly Nordic electricity spot prices (system
          price) for the period from May 4, 1992 to December 31,
          2000~\protect\cite{NordPool}. The total number of data
          points is somewhat larger then $75\,000$. The inset shows
          the spot prices for the first week of year 2000 staring on
          Monday January 3. From the inset one should observe the
          daily price trends.}
        \label{Fig:Price}
    \end{center}
\end{figure}

\begin{figure}[tbp]
    \begin{center}
        \leavevmode
         \includegraphics[clip,scale=0.5]{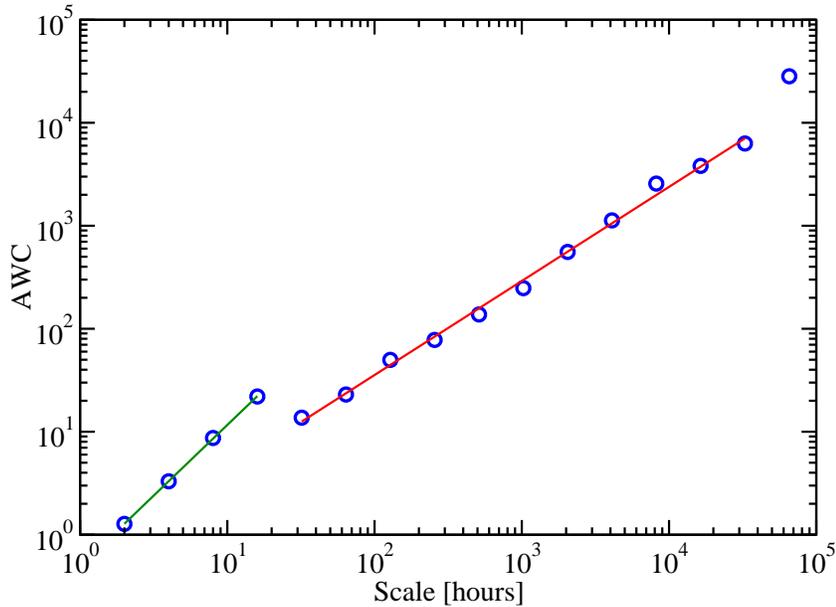}
        \caption{The AWC spectrum, $W[p](a)$ vs. scale $a$, for the
          hourly Nordic electricity spot data presented in
          Fig.~\protect\ref{Fig:Price}. A cross-over at $a_\times\sim
          24\,h$ is easily observed in the $W[p](a)$-spectrum.  The
          scaling region $a>a_\times$ corresponds to a Hurst exponent
          of $H=0.41 \pm 0.02$ where the uncertainty is a pure
          regression error. The slope of the spectrum for
          $a<a_\times$ seems to indicate a persistent behavior
          ($H>0.5$). The wavelet used in obtaining these results was
          of the Daubechies type (DAUB24). }
        \label{Fig:AWC}
    \end{center}
\end{figure}

In Fig.~\ref{Fig:AWC} we present the AWC-spectrum, $W[p](a)$, for the
Nordic electricity spot price time series depicted in
Fig.~\ref{Fig:Price}.  Probably the most striking feature of this
spectrum is the nice and large scaling regime for scales bigger then
one day ($a>24\,h$).  Its size extends over three orders of magnitude
in time ranging from days to several years.  The Hurst exponent
characterizing this scaling region is found by a regression fit to the
functional form given by Eq.~(\ref{Eq:AWC-scaling}).  Such a fit,
indicated by a solid line in Fig.~\ref{Fig:AWC}, results in a Hurst
exponent of 
\begin{eqnarray}
  \label{eq:exponent}
  H=0.41 \pm 0.02 ,  
\end{eqnarray}
where the error bar is a pure regression error.  This value for the
Hurst exponent means that the spot prices process is {\em
  anti-persistent} (anti-correlated) or equivalently {\em
  mean-reverting}. Another way of saying the same is that a price
drop, say, in the past is more likely in the future to be followed by
an increase in the price then by another price drop.

Due to the entrance of Sweden and Finland into the spot market in the
years of 1995 and 1996, it is interesting to see if this did affect
the statistical properties of the spot market in any significant way.
To study this, we have compared the Hurst exponents for the two
periods 1992--1995 and 1996--2000. We found that within the error
bars, the Hurst exponents for these two periods could not be
distinguished, even though the numerical value of the exponent was
decreased somewhat from the first to the second period.

It is also interesting to notice that the value measured for the Hurst
exponent of the Nordic spot market, Eq. (\ref{eq:exponent}), is more
or less the same as the one reported earlier for the Californian
market.  For this latter market Weron and
Przyby{\l}owicz~\cite{Weron-2000-A} found $H=0.42-0.43$.  For the
moment it is too early to say if the reasons for these values being so
similar is just a coincidence or there exists more fundamental reasons
for it.

The size, and quality, of the scaling regime for $a>24\, h$ we find
somewhat surprising. First, the data contains of both daily as well as
seasonal trends, non of which have been removed from the analyzed data
set. Since we are using an analyzing method that decouples scales, the
daily trend should not influence the (day-to-day) scales considered.
However, the seasonal trends, typically with highest prices during
winter time, could in principle affect the scaling regime. This trend
seems, however, to have little or no effect on the spectrum in
practice~\cite{reason}.  Hence, we can for instance by studying the
week-to-week price changes get information about the year-to-year
price changes simply by a rescaling of the problem according to
Eq.~(\ref{Eq:Self-affine}) with the value for the Hurst exponent given
by Eq.~(\ref{eq:exponent}) --- {\it i.e.} we have scale invariance!

We just saw that the Nordic electricity spot prices on a time scale
bigger then one day is well approximated by an anti-persistent
self-affine walk.  Such anti-correlations are rather atypical for
financial time series.  For instance for liquid stock indices one
typically finds them to be uncorrelated after a rather short time
period~\cite{Book:Bouchaud-2000,Book:Mantegna-2000}.  Anything else
would give rise to arbitrage opportunities due to the correlated (or
anti-correlated) increments~\cite{ourpaper}.  By exploiting these
correlations by for instance adapting an investment strategy that
takes them into consideration, they will most likely vanish. That this
apparently does not happen for the electricity spot market we believe
is related to the fact that electricity, for the moment, can not be
stored efficiently.  Even if we knew for sure that the spot price
would increase tomorrow, we could not buy electricity today and sell
it with profit tomorrow simply because we do not have a way of storing
it in the mean time in an economical way.  Therefore, there seems at
present to be no profit opportunity created by the anti-correlation in
the spot price increments.  However, if technological advances in the
future will allow for an efficient storage of electricity (at moderate
cost), the situation will probably be quite different.  Then profit
opportunities will exist, and we expect anti-correlations to disappear
for time scales where efficient storage of electricity is feasible.

Another striking feature of the AWC-spectrum shown in
Fig.~\ref{Fig:AWC} is the sharp cross-over taking place at roughly one
day, {\it i.e.}, at a scale $a=a_\times=24\,h$. At scales bigger then
this, we just saw that we had anti-correlation. However, at smaller
scales, it seems more reasonable to assume that the price increments
are correlated.  If we temporary assume a scaling regime for scales
$a<a_\times$ (that we will not insist on), one from a regression fit
obtains $H=0.87 \pm 0.02$.  Hence, it should be clear that the spot
price time series behaves differently on an intra-day scale then on a
day-to-day scale (or bigger). In fact, such an observation can be done
directly from Fig.~\ref{Fig:Price} by visual inspection.

Finally, we would like to give some explicit examples showing the need
of an analyzing method that decouples scales when handling multi-scale
time series.  This we will do by applying the classic rescaled-range
(or $R/S$) and the popular Fourier analyzing technique to our data
set~\cite{Book:Feder-1988}. Notice that none of these two methods do
decouple the scales involved. In the $R/S$-analysis, one calculates
the range, $R(w)$, defined as the average vertical distance between
the maximum and minimum point of the cumulative signal calculated over
a window of size $w$, divided by the (average) standard deviation,
$S(w)$, of the original signal taken over the same window size.  It
can be shown that the ratio $R/S$ of a self-affine series should scale
according to $w^H$.  The Fourier method, on the other hand, is
performed by calculating the power spectrum of the time-series,
$P(\omega)$, where $\omega=2\pi/\nu$ denotes the angular frequency
with $\nu$ being the frequency. For a self-affine time series, this
quantity should scale as $P(\omega)\sim \omega^{-2H-1}$.  The
interested reader is referred to Ref.~\cite{Book:Feder-1988} for a
detailed introduction to both these two methods.

\begin{figure}[tbp]
    \begin{center}
      \leavevmode \includegraphics[scale=0.5]{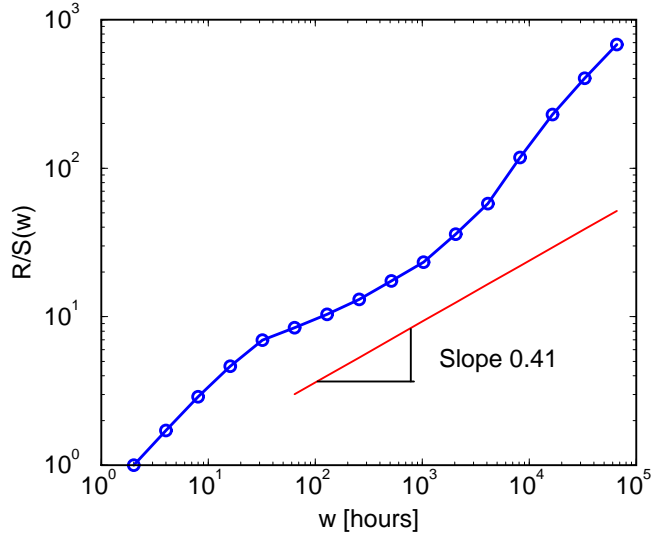}
       \caption{The $R/S$-analysis of the hourly Nordic electricity spot data
         presented in Fig.~\protect\ref{Fig:Price}. Notice that no
         scaling regime can be defined for window sizes bigger then
         one day (see text for explanation). The solid line
         corresponds to $w^H$ with $H=0.41$ as measured by the
         AWC-method.}
        \label{Fig:RS}
    \end{center}
\end{figure}

The result of a $R/S$ analysis of the NordPool spot data is presented
in Fig.~\ref{Fig:RS}. We observe that only for a small window size (or
scale) corresponding to a day or less, do we have a signature of a
scaling regime, but, as above, it is too small to be called so. This
region is what corresponds to what we found for the intra-day price
increments ($a<24\,h$) in our wavelet analysis, and the corresponding
exponents obtained from the two methods are also found to be similar.
However, when we focus at bigger scales, we notice, according to the
$R/S$ technique, that there is no scaling region since the
$R/S$-function is curved in this region. From the wavelet analysis,
however, we have seen that there should be a rather robust scaling
regime for $a>a_\times$.  The solid line in Fig.~\ref{Fig:RS}
indicates the slope corresponding to the Hurst exponent previously
measured by the wavelet technique and given by
Eq.~(\ref{eq:exponent}).  Such a regime can hardly be defined for the
$R/S$-function in this region.  The reason why the scaling regime is
not found by the rescaled-range method is that the calculation of the
$R/S$-function for large window sizes is influenced substantially by
smaller window sizes.  This is particularly seen at the cross-over;
Just above the cross-over, the rescaled-range signal will be dominate
by the behavior below the cross-over, resulting in a {\em smear-out}
in this region. One observes from Fig.~\ref{Fig:RS} that this
cross-over transition hampers the scaling behavior for several orders
of magnitude in scale after the cross-over. In fact for this
particular case, no scaling is possible to observe at all for scales
ranging from the cross-over and up to the maximum scale studied.
Thus, we may conclude that the $R/S$-method is not well suited for
analyzing time-series that possess multiple scale behavior.  This is
the reason why Weron and Przyby{\l}owicz~\cite{Weron-2000-A} analyzed,
by the rescaled-range method, the daily averaged returns instead of
the hourly price increments themselves.  We have also checked that
analyzing the daily averaged returns of our data set by the
$R/S$-method produces a scaling regime of a Hurst exponent consistent
with the one previously found by the wavelet technique.

\begin{figure}[tbp]
    \begin{center}
        \leavevmode
         \includegraphics[scale=0.5,angle=-90]{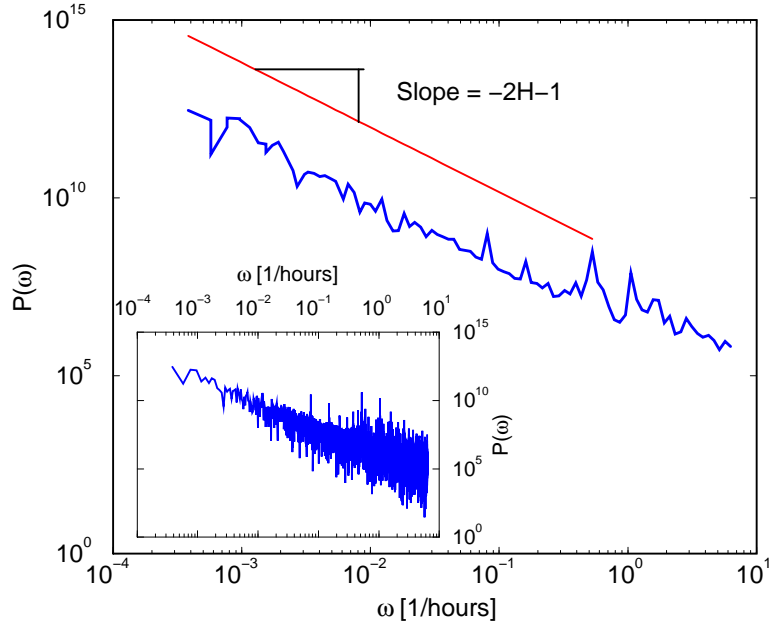}
       \caption{The Fourier-analysis of the hourly Nordic electricity spot data
         presented in Fig.~\protect\ref{Fig:Price}. The inset shows
         the {\em raw} power spectrum, $P(\omega)$ vs. angular
         frequency $\omega$ for the spot price data. In order to
         reduce the noise of the raw power spectrum, we have applied
         the log-binning technique to it, resulting in the power
         spectrum shown in the main figure. The solid line of this
         figure corresponds to $\omega^{-2H-1}$ with the value of $H$
         obtained from the AWC-technique
         (Eq.~(\protect\ref{eq:exponent})). The pronounced cross-over
         seen in Fig.~\protect\ref{Fig:AWC} corresponds to an angular
         frequency of $\omega_\times=2\pi/a_\times \sim 0.26\,
         h^{-1}$, but such a cross-over is not easily located in the
         power spectrum. A direct regression fit to the log-binned
         power spectrum for $\omega<\omega_\times$ results in
         $H=0.42\pm 0.04$. }
        \label{Fig:Fourier}
    \end{center}
\end{figure}

In the inset to Fig.~\ref{Fig:Fourier} we present the power spectrum
$P(\omega)$ vs. (angular) frequency $\omega$ for our time series. It
is observed, as is a rather typical case, that the raw power spectrum
is somewhat noisy, and that the cross-over that we know should be
there is not so easy to locate. In order to reduce the noise level, it
is rather common to use a so-called log-binning technique. This
amounts to using a gliding window of a size that increases
logarithmically with frequency, and to take the average of the power
spectrum within this window.  Strictly speaking, this is not a
rigorous approach, but if the window size is not too large it does not
seem to have any practical consequence for the estimated scaling
exponents.  The results of such a procedure is shown in the main part
of Fig.~\ref{Fig:Fourier}, and one observes a dramatic reduction in
the noise.  The solid line in this figure represents the
$\omega^{-2H-1}$ behavior where for the Hurst exponent we have used
that of Eq.~(\ref{eq:exponent}) (the wavelet result). For the Fourier
method, in contrast to the rescaled-range method, there is a
reasonable agreement with the wavelet method when the hourly price
data are used.  If we measure the Hurst exponent from the (log-binned)
power spectrum we find $H=0.42\pm 0.04$ that should be compared with
the value $H=0.41 \pm 0.02$ that we obtained by the AWC-method. As
before, the error bars are pure regression errors, and the real error
is of course somewhat bigger.

Notice, however, that even after log-binning the cross-over in the
power spectrum expected to be located at
$\omega_\times=2\pi/a_\times=0.26\, h^{-1}$, is not that sharp or well
pronounced, even though there seems to be some indication of it. The
two peaks in the power spectrum at approximately $\omega\simeq 0.5\,
h^{-1}$ and $1\, h^{-1}$, corresponding in scale ($a=2\pi/\omega$) to
$a\simeq 12\, h$ and $6\, h$ respectively, are caused by the intra-day
price increase due to the consumption pattern with high consumption in
the morning and afternoon (see the inset to Fig.~\ref{Fig:Price}).
This effect was not caught by the wavelet technique used in this
paper.  We believe that the reason for this is the use of the discrete
wavelet transform which does not include the scales $6\,h$ or $12\,
h$.  However, such enhanced correlations, we suspect, will be caught
if the {\em continuous} wavelet transform is used instead of a
discrete one~\cite{Work-in-progree}.

\section{Conclusions}

We have analyzed the hourly logged spot electricity prices from the
Nordic electricity power market over a eight and a half years period
starting in mid 1992.  It is found that the spot price process is an
anti-persistent self-affine walk (mean-reverting stochastic process)
over at least three orders of magnitude in time ranging from days to
years. Hence the increments in the spot price process are
anti-correlated over the same time scales. The Hurst exponent that
characterize this scaling behavior was measured by a wavelet technique
and found to be $H=0.41 \pm 0.02$, where the error bar is a pure
regression error.  We also commented on other classical methods of
measuring such exponents and stressed and showed that it is important
to use methods that decouple scales when multi-scale time series are
being analyzed.

\section*{Acknowledgements}

We would like to thank SKM Kraft AS for providing us with the data
analyzed in the present paper.  The author would also like to express
his gratitude to H.\ Gr{\o}nlie and C.\ Land{\aa}s for numerous
discussion related to the Nordic Power market.  The critical reading
of the manuscript and the helpful comments provided by Kim Sneppen is
also highly acknowledged.

\bibliographystyle{prsty}  
\bibliography{econophysics,books,publication_list,self-affine,paper2001-1-bibtex}

\begin{thebibliography}{10}
\bibitem{comment}
Here we could also have used another norm $W_\nu[p](a) = \left< \left|{\mathcal
  W}(a,b)\right|^\nu\right>^{1/\nu}_b$ with $\nu\neq 1$, but we will not
  consider this option here. In practice the results show little sensitivity to
  the choice of $\nu$ (for moderate values).
\end{thebibliography}



\newpage

\end{document}